\documentclass[a4paper]{JHEP3}
\usepackage{epsfig}

\title{Higher order QED corrections to muon decay spectrum}

\author{Andrej Arbuzov\thanks{The work was performed in
Department of Physics, University of Alberta, Canada}\\
BLTPh, Joint Institute for Nuclear Research, Joliot Curie 6, Dubna, 141980, Russia \\
E-mail: \email{arbuzov@thsun1.jinr.ru}
}

\abstract{QED radiative corrections to
polarized muon decay spectrum are considered.
Leading and next--to--leading logarithmic approximations are used.
Exponentiation of soft radiation is discussed.
The present theoretical uncertainty of the spectrum description
is estimated.}

\keywords{smo.lde.emp}

\preprint{hep-ph/0206036 \\ Alberta-Thy-10-02}

\begin{document}

\def\TWIST{$\mathcal{TWIST}\ $}
\def\DD{{\mathcal D}}
\def\NS{\mathrm{NS}}
\def\SS{\mathrm{S}}
\def\LL{\mathrm{LL}}
\def\NLL{\mathrm{NLL}}
\def\MSbar{$\overline{\mathrm{MS}}\ $}
\def\Li#1#2{{\mathrm{Li}}_{#1}\left(#2\right)}
\def\Sot#1{{\mathrm{S}}_{1,2}\left(#1\right)}
\def\ba{\begin{eqnarray}}
\def\ea{\end{eqnarray}}
\def\dd{{\mathrm d}}
\def\la{\mathrel{\mathpalette\fun <}}
\def\ga{\mathrel{\mathpalette\fun >}}
\def\fun#1#2{\lower3.6pt\vbox{\baselineskip0pt\lineskip.9pt
  \ialign{$\mathsurround=0pt#1\hfil##\hfil$\crcr#2\crcr\sim\crcr}}}
\def\order#1{{\mathcal O}\left(#1\right)}

\section{\label{Sec:Int}Introduction}

Accurate measurements of the muon properties were providing substantial
information for the development of the elementary particle physics during
many years. Nowadays precision experiments with muons serve as one of the
basements of the Standard Model (SM) and give a possibility to look for
{\em new physics}~\cite{Fetscher:2000th,Kuno:2001jp}.

In this paper we discuss the present theoretical precision
of the polarized muon decay spectrum description.
The study is motivated by the experiment
\TWIST\cite{Rodning:2001js,Quraan:2000vq}, which
is currently running at Canada's National Laboratory TRIUMF.
The experiment is going to measure the spectrum with the accuracy level
of about $1\cdot 10^{-4}$.
That will make a serious test of the space--time structure
of the weak interaction. The experiment is able to put stringent limits
on a bunch of parameters in models beyond SM, {\it e.g.}, on the mass
and the mixing angle of a possible right--handed $W$-boson.
To confront the experimental results with SM,
adequately accurate theoretical predictions should
be provided. This necessarily requires to
calculate radiative corrections within the perturbative
Quantum Electrodynamics (QED). In this paper
we will concentrate on the effect of higher order leading
and next--to--leading logarithmic corrections.

\section{\label{Sec:Pre}Preliminaries}

Within the Standard Model, the differential
distribution of electrons (averaged over electron spin states)
in the polarized muon decay reads
\ba \label{general}
&& \frac{\dd^2\Gamma^{\mu^{\mp}\to
e^{\mp}\nu\bar{\nu}}}{\dd x\dd c}
= \Gamma_0 \left( F(x) \pm cP_{\mu} G(x) \right), \qquad
\Gamma_0 = \frac{G_F^2 m_\mu^5}{192\pi^3}\, ,
\nonumber \\ &&
c = \cos\theta, \qquad
x = \frac{2m_{\mu}E_e}{m_\mu^2+m_e^2}, \qquad
x_0 \leq x \leq 1, \qquad
x_0 = \frac{2m_{\mu}m_e}{m_\mu^2+m_e^2},
\ea
where
$m_\mu$ and $m_e$ are the muon and electron masses;
$G_F$ is the Fermi coupling constant;
$\theta$ is the angle between the muon polarization vector $\vec{P}_{\mu}$
and the electron (or positron) momentum;
$E_e$ and $x$ are the energy and the energy fraction of $e^{\pm}$.
Here we adopt the definition of the Fermi coupling constant following
Ref.~\cite{vanRitbergen:2000fi}.
Within the Standard Model, the muon decay
happens due to the weak interaction of leptons and $W$-bosons. The Fermi
model corresponds to the limiting case of the infinite $W$-boson mass.
If $G_F$ is defined according to Refs.~\cite{Groom:2000in,Marciano:1988vm},
the first order effect in the muon and $W$-boson mass ratio gives
\ba
\Gamma_0 \longrightarrow \Gamma_0
\biggl( 1 + \frac{3}{5}\;\frac{m_\mu^2}{m_W^2} \biggr).
\ea

Functions $F(x)$ and $G(x)$ describe the isotropic and anisotropic
parts of the spectrum, respectively. Within perturbative QED, they
can be expanded in series in the fine structure constant $\alpha$:
\ba
F(x) = f_{\mathrm{Born}}(x) + \frac{\alpha}{2\pi}f_1(x)
+ \biggl(\frac{\alpha}{2\pi}\biggr)^2f_2(x)
+ \biggl(\frac{\alpha}{2\pi}\biggr)^3f_3(x)
+ \order{\alpha^4},
\ea
and in the same way for $G(x)$. The Born--level functions
$f_{\mathrm{Born}}$ and $g_{\mathrm{Born}}$ are well known,
including small terms suppressed by the $m_e/m_\mu$ mass
ratio~\cite{Fetscher:2000th}:
\ba
f_{\mathrm{Born}}(x) &=& 6x\biggl(1+\frac{m_e^2}{m_\mu^2}\biggr)^4
\sqrt{1-\frac{m_e^2}{E_e^2}}\; \biggl[ x(1-x)
+ \frac{2}{9}\rho ( 4x^2 - 3x - x_0^2 )
+ \eta  x_0(1-x) \biggr],
\nonumber \\
g_{\mathrm{Born}}(x) &=& -2x^2 \xi \biggl(1+\frac{m_e^2}{m_\mu^2}\biggr)^4
\biggl(1-\frac{m_e^2}{E_e^2}\biggr) \biggl[ 1 - x
+ \frac{2}{3}\delta \biggl(4x - 3 - \frac{m_e}{m_\mu}x_0\biggr) \biggr],
\ea
where $\rho$, $\eta$, $\xi$, and $\delta$ are the so--called Michel
parameters~\cite{mich1,mich2,mich3}, which in the Standard Model
take the following values:
$\rho=3/4$, $\eta=0$, $\xi=1$, and $\delta=3/4$.
By fitting the values of the parameters from the experimental data
and comparing them with the SM predictions, the \TWIST experiment is going
to look for effects of non--standard interactions.

The first order corrections $f_1(x)$ and $g_1(x)$ are also known
with the exact account of the dependence on the electron
mass~\cite{Behrends:1956mb,Kinoshita:1959ru,Arbuzov:2001ui}.
Starting from $\order{\alpha}$, radiative corrections to
the electron spectrum contain so--called mass singularities
in the form of the large logarithm
$L\equiv\ln(m^2_\mu/m_e^2)\approx 10.66$.
As demonstrated in Ref.~\cite{Arbuzov:2002pp}
(see also Table~\ref{table:1} below), the terms of the order
$\order{\alpha L}$ give the bulk of the first order correction.
This is our reason to look first for higher order terms enhanced by the
large logarithm. These enhanced terms (excluding the
ones related to the running of the QED coupling constant)
cancel out in the expression for the total muon decay width at any order
in $\alpha$ in accord with the Kinoshita--Lee--Nauenberg
theorem~\cite{Kinoshita:1962ur,Lee:1964is}.

The second order corrections to the total muon decay width were
calculated in Ref.~\cite{vanRitbergen:1998yd}.
At this order for the differential decay spectrum, we know
only the leading logarithmic (LL) corrections~\cite{Arbuzov:2002pp}
and the isotropic part in the next--to--leading logarithmic (NLL)
approximation~\cite{Arbuzov:2002cn}. The corresponding anisotropic part
will be given below. The third order LL corrections
will be presented here as well.

\section{\label{Sec:FFA}The Fragmentation Function Approach}

First, I will describe briefly the application of the renormalization
group method to the calculation of the leading and next--to--leading
radiative corrections to the polarized muon decay spectrum.
For a detailed foundation of the procedure and
an extended discussion see Ref.~\cite{Arbuzov:2002cn}.

The factorizations theorems, proved for QCD~\cite{Ellis:qj},
can be easily translated to QED. In particular, by means of
factorization, one can represent the differential
spectrum of electrons as a convolution:
\ba \label{master}
\frac{\dd^2\Gamma}{\dd x\dd c}(x,c,m_\mu,m_e) =
\sum\limits_{j=e,\gamma}^{}
\int\limits_{x}^{1}\frac{\dd z}{z}\;
\frac{\dd^2\hat{\Gamma}_j}{\dd z\dd c}(z,c,m_\mu,\mu_f)
\DD_j\biggl(\frac{x}{z},\mu_f,m_e\biggr),
\ea
where $\dd^2\hat{\Gamma}_j/(\dd z\dd c)$
is the differential distribution a muon decay with production of
a massless electron $(j=e)$ or a photon $(j=\gamma)$
with energy fraction $z$ ~$(z=2E_j/m_\mu$, where $E_j$ is
the energy of the relevant particle).
To subtract collinear singularities from the
differential distributions, we will use here
the \MSbar factorization scheme~\cite{Bardeen:1978yd} with the
scale $\mu_f$. The fragmentation function
$\DD_j(x/z,\mu_f,m_e)$ describes the conversion
of a massless parton $j$ into a massive {\em physical} electron.

The spectrum of the massless
parton can be expanded in a perturbative series:
\ba \label{dhGj}
\frac{1}{\Gamma_0}\frac{\dd^2\hat{\Gamma}_j}{\dd z\dd c}(z,c,m_\mu,\mu_f)
= \hat{A}^{(0)}_j(z,c)
+ \frac{\bar{\alpha}(\mu_f)}{2\pi}\hat{A}^{(1)}_j(z,c)
+ \biggl(\frac{\bar{\alpha}(\mu_f)}{2\pi}\biggr)^2\hat{A}^{(2)}_j(z,c)
+ \order{\alpha^3},
\ea
where $\bar{\alpha}(\mu_f)$ is the
\MSbar renormalized coupling constant, which will be converted
further into the traditional QED on--shell coupling constant
$\alpha \approx 1/137.036$.
The lowest order spectrum of massless electrons is defined by
\ba
\hat{A}^{(0)}_e(z,c) = f_0(z) \pm cP_{\mu} g_0(z), \qquad
f_0(z) = z^2(3 - 2z), \qquad
g_0(z) = z^2(1 - 2z).
\ea
Here and in what follows, the sign before $c$ should be chosen
according to the charge of the decaying muon (plus for
$\mu^-$ decay and {\it vice versa}).
The $\order{\alpha}$ correction to the massless electron spectrum reads
\ba \label{hAe}
\hat{A}^{(1)}_e(z,c) &=& \hat{f}^{(1)}_e(z)
\pm cP_{\mu} \hat{g}^{(1)}_e(z),
\\ \label{hge}
\hat{g}^{(1)}_e(z) &=& \biggl( 2z^2(1-2z)\ln\frac{1-z}{z}
- \frac{1}{6} - 4z^2 + \frac{8}{3}z^3 \biggr)\ln\frac{m_\mu^2}{\mu_f^2}
\nonumber \\
&+& 2z^2(1-2z)\biggl( \ln^2(1-z) - 4\Li{2}{1-z}
- \ln z\ln(1-z)  - 2\ln^2z \biggr)
\nonumber \\
&+& \biggl( \frac{11}{3} - \frac{4}{3z} - 6z
- \frac{17}{3}z^2 + \frac{34}{3}z^3 \biggr)\ln(1-z)
+ \biggl( - \frac{1}{3} - 6z^2 - 6z^3 \biggr)\ln z
\nonumber \\
&-& \frac{7}{6} + 3z + \frac{7}{6}z^2 + 3z^3.
\ea
For the auxiliary photon spectrum with collinear singularities
subtracted according to the \MSbar prescription, we have
\ba \label{hAg}
\hat{A}^{(0)}_{\gamma}(z,c) &=& 0, \qquad
\hat{A}^{(1)}_{\gamma}(z,c) = \hat{f}^{(1)}_{\gamma}(z)
\pm cP_{\mu} \hat{g}^{(1)}_{\gamma}(z),
\\ \label{hgg}
\hat{g}^{(1)}_{\gamma}(z) &=&
\biggl( \frac{1}{3} - \frac{1}{3z} - \frac{2}{3}z^2
+ \frac{2}{3}z^3 \biggr)
\biggl(\ln\frac{m_\mu^2}{\mu_f^2} + \ln(1-z)\biggr)
\nonumber \\
&+& \biggl( \frac{2}{3} - \frac{2}{3z} \biggr)\ln z
- \frac{2}{3} + \frac{2}{3z} + \frac{11}{12}z
- \frac{2}{3}z^2 - \frac{1}{4}z^3.
\ea
The isotropic parts of the first order corrections to the
auxiliary massless parton distributions,
$\hat{f}^{(1)}_e(z)$ and $\hat{f}^{(1)}_{\gamma}(z)$,
are given by Eqs.~(7,8) in Ref.~\cite{Arbuzov:2002cn}.
By the choice of the factorization parameter value,
$\mu_f\sim m_\mu$, we avoid calculation of the unknown
functions $\hat{A}^{(2)}_j$, since then they can not give
rise to any LL or NLL correction.

The fragmentation functions $\DD_j$ can be obtained by
solving the Dokshitzer--Gribov--Lipatov--Altarelli--Parisi (DGLAP)
evolution equations for QED,
\ba
\label{DGLAP}
\frac{\dd\DD_i(x,\mu_f,m_e)}{\dd \ln\mu_f^2} = \sum\limits_{j=e,\gamma}
\int\limits_{x}^{1}\frac{\dd z}{z}\;
P_{ji}(z,\bar{\alpha}(\mu_f))
\DD_j\biggl(\frac{x}{z}\, ,\mu_f,m_e\biggr),
\ea
where $P_{ji}$ are perturbative splitting functions,
\ba
P_{ji}(z,\bar{\alpha}(\mu_f)) =
\frac{\bar{\alpha}(\mu_f)}{2\pi}P_{ji}^{(0)}(z)
+ \biggl(\frac{\bar{\alpha}(\mu_f)}{2\pi}\biggr)^2P^{(1)}_{ji}(z)
+ \order{\alpha^3}.
\ea
The initial conditions, which are required to solve the DGLAP
equations by iterations, can be borrowed from QCD
studies~\cite{Mele:1990cw}:
\ba \label{Deini}
\DD^{\mathrm{ini}}_e(x,\mu_0,m_e) &=& \delta(1-x)
+ \frac{\bar{\alpha}(\mu_0)}{2\pi} d_1(x,\mu_0,m_e) + \order{\alpha^2},
\nonumber \\
d_1(x,\mu_0,m_e) &=& \biggl[ \frac{1+x^2}{1-x}
\biggl( \ln\frac{\mu_0^2}{m_e^2} - 2\ln(1-x) - 1 \biggr)
\biggr]_+,
\\ \label{Dgini}
\DD^{\mathrm{ini}}_\gamma(x,\mu_0,m_e) &=& \frac{\bar{\alpha}(\mu_0)}{2\pi}
\ln\frac{\mu_0^2}{m_e^2}
P^{(0)}_{\gamma e}(x) + \order{\alpha^2}.
\ea
The relevant lowest order splitting functions are
\ba
P_{ee}^{(0)}(x)= \biggl[\frac{1+x^2}{1-x}\biggr]_+,
\qquad
P^{(0)}_{e\gamma}(x) = \frac{1 + (1-x)^2}{x}\, ,
\qquad
P^{(0)}_{\gamma e}(x) = x^2 + (1-x)^2.
\ea
The {\em plus} prescription works as usually:
\ba
&& \int\limits_{x_{\mathrm{min}}}^1\dd x\; [V(x)]_+W(x) =
\int\limits_{0}^1\dd x\; V(x) [W(x)\Theta(x-x_{\mathrm{min}})-W(1)],
\\ \nonumber &&
\Theta(x) =
\left\{\begin{array}{l} 1 \quad {\mathrm{for}} \quad x\geq 0 \\
0 \quad {\mathrm{for}} \quad x < 0
\end{array}\right. .
\ea

In a measurement of the muon decay spectrum,
events with more than one electron in the final state require a special
treatment.
Starting from the second order in $\alpha$, we have a certain
contribution due to emission of real and virtual $e^+e^-$ pairs.
Presumably a Monte Carlo event generator is needed to simulate
the process of muon decay with pair production.
Nevertheless, we will calculate the corresponding effect
under a simple assumption, that an event with two electrons in the final state
is treated as a pair of simultaneous muon decays. In order
to have the possibility to drop the pair contributions (if they
are taken into account in a Monte Carlo program), we decompose our results
according to the classes of the corresponding Feynman diagrams
in the same way as in Ref.~\cite{Berends:1987ab}.
Moreover, the decomposition will help us to demonstrate the
cancellation of the mass singularities in the integrated decay width.
Our results for pure pair corrections can serve further also as
a benchmark for the Monte Carlo program.
The next--to--leading electron splitting function
can be decomposed into four parts:
\ba
P^{(1)}_{ee}(x) = P^{(1,\gamma)}_{ee}(x)
+ P^{(1,\NS)}_{ee}(x)
+ P^{(1,\SS)}_{ee}(x)
+ P^{(1,\mathrm{int})}_{ee}(x),
\ea
where $P^{(1,\gamma)}_{ee}(x)$ is provided by the set of Feynman
diagrams with pure photonic corrections;
$P^{(1,\NS)}_{ee}(x)$ is related to the corrections due to non--singlet
real and virtual pairs;
$P^{(1,\SS)}_{ee}(x)$ stands for the singlet pair production
contribution; and $P^{(1,\mathrm{int})}_{ee}(x)$ describes
the interference of real singlet and non--singlet pairs.
By extracting the appropriate color structures from the known
QCD results~\cite{Curci:1980uw,Floratos:1981hs,Furmanski:1980cm,Ellis:1996nn},
the explicit expressions for these functions were
given in Ref.~\cite{Arbuzov:2002cn}.
\label{def:sns}
Here and in what follows, in the language of Feynman diagrams,
the situation when the registered electron is connected by a solid
fermion line with the genuine electroweak decay vertex is
called {\em non--singlet}. The case, when the observed electron
belongs to a pair produced via a virtual photon, is called {\em singlet}.

The relation between the \MSbar and the on--shell coupling constants
reads~\cite{vanRitbergen:2000fi}
\ba
\bar{\alpha}(\mu_f) = \alpha + \frac{\alpha^2}{3\pi}\ln\frac{\mu_f^2}{m_e^2}
+ \frac{\alpha^3}{4\pi^2} \ln\frac{\mu_f^2}{m_e^2}
+ \frac{15\alpha^3}{16\pi^2} + \order{\alpha^4}.
\ea

It is convenient to choose the renormalization scale to be
\ba
\mu_0 = m_e.
\ea

Now we have everything for solving the DGLAP equations~(\ref{DGLAP}).
Using iterations for the electron fragmentation function decomposed
into four parts, we get
\ba \label{De}
\DD_e(x,\mu_f,m_e) &=& \DD_e^{(\gamma)}(x)
+ \DD_e^{(\NS)}(x)
+ \DD_e^{(\SS)}(x)
+ \DD_e^{(\mathrm{int})}(x),
\\ \label{Deg}
\DD_e^{(\gamma)}(x) &=& \delta(1-x)
+ \frac{\alpha}{2\pi}d_1(x,\mu_0,m_e)
+ \frac{\alpha}{2\pi}L_{f}P^{(0)}_{ee}(x)
\nonumber \\
&+& \biggl(\frac{\alpha}{2\pi}\biggr)^2
\biggl(\frac{1}{2}L_{f}^2P^{(0)}_{ee}\otimes P^{(0)}_{ee}(x)
+ L_{f}P^{(0)}_{ee}\otimes d_1(x,\mu_0,m_e)
+ L_{f}P^{(1,\gamma)}_{ee}(x) \biggr)
\nonumber \\
&+& \biggl(\frac{\alpha}{2\pi}\biggr)^3
\frac{1}{6}L_{f}^3P^{(0)}_{ee}\otimes P^{(0)}_{ee}\otimes P^{(0)}_{ee}(x),
\\ \label{DeNS}
\DD_e^{(\NS)}(x) &=&
\biggl(\frac{\alpha}{2\pi}\biggr)^2
\biggl( \frac{1}{3}L_{f}^2P^{(0)}_{ee}(x)
+ L_{f}P^{(1,\NS)}_{ee}(x) \biggr)
\nonumber \\
&+& \biggl(\frac{\alpha}{2\pi}\biggr)^3L_{f}^3
\biggl(\frac{1}{3}P^{(0)}_{ee}\otimes P^{(0)}_{ee}(x)
+ \frac{4}{27}P^{(0)}_{ee}(x) \biggr),
\\ \label{DeS}
\DD_e^{(\SS)}(x) &=&
\biggl(\frac{\alpha}{2\pi}\biggr)^2
\biggl( \frac{1}{2}L_{f}^2P^{(0)}_{e\gamma}\otimes P^{(0)}_{\gamma e}(x)
+ L_{f}P^{(1,\SS)}_{ee}(x) \biggr)
\nonumber \\
&+& \biggl(\frac{\alpha}{2\pi}\biggr)^3L_{f}^3
\biggl( \frac{1}{3}P^{(0)}_{e\gamma}\otimes P^{(0)}_{\gamma e}
\otimes P^{(0)}_{ee}(x)
- \frac{1}{9}P^{(0)}_{e\gamma}\otimes P^{(0)}_{\gamma e}(x) \biggr),
\\ \label{Deint}
\DD_e^{(\mathrm{int})}(x) &=&
\biggl(\frac{\alpha}{2\pi}\biggr)^2L_{f}P^{(1,\mathrm{int})}_{ee}(x),
\qquad
L_f \equiv \ln\frac{\mu_f^2}{m_e^2}\, ,
\ea
where we systematically omitted terms of the following orders:
$\order{\alpha^2L_f^0}$, $\order{\alpha^3L_f^2}$, $\order{\alpha^4}$,
and higher.
The photon fragmentation function at the lowest order,
\ba \label{Dg}
\DD_\gamma(x,\mu_f,m_e) = \frac{\alpha}{2\pi}
L_{f}P^{(0)}_{e\gamma}(x) + \order{\alpha^2},
\ea
is sufficient for our purposes.
The convolution operation is defined in the standard way:
\ba
\label{conv}
A\otimes B(x) = \int\limits^1_0\dd z
\int\limits^1_0\dd z'\; \delta(x-zz')A(z)B(z')
=\int\limits^1_x\frac{\dd z}{z}\; A(z)B\biggl(\frac{x}{z}\biggr).
\ea

The leading logarithmic terms for the QED
fragmentation function are known up to the fifth order in
$\alpha$~\cite{Arbuzov:1999cq}. But, as will be seen from the numerical
results, keeping contributions up to the third order is enough
for the moment.

The explicit expressions for the functions, which appear in the
$\order{\alpha^2}$ terms of the function $\DD_e$,
are given in Ref.~\cite{Arbuzov:2002cn}. In the third order we have
two more functions~\cite{Skrzypek:1992vk}:
\ba \label{P3}
&& P^{(0)}_{ee}\otimes P^{(0)}_{ee}\otimes P^{(0)}_{ee}(x) =
\delta(1-x)\biggl( 16\zeta_3 - \frac{81}{4} \biggr)
+ \biggl[ 24\frac{1+x^2}{1-x}\biggl( \frac{1}{2}\ln^2(1-x)
+ \frac{3}{4}\ln(1-x)
\nonumber \\ && \qquad
+ \frac{9}{32} - \frac{1}{2}\zeta_2 \biggr) \biggr]_+
+ 24\frac{1+x^2}{1-x}\biggl( \frac{1}{12}\ln^2x
- \frac{1}{2}\ln{x}\ln(1-x)
- \frac{3}{8}\ln{x} \biggr)
\nonumber \\ && \qquad
+ 6(1+x)\ln{x}\ln(1-x)
- 12(1-x)\ln(1-x)
+ \frac{3}{2}(5-3x)\ln{x}
- 3(1-x)
\nonumber \\ && \qquad
- \frac{3}{2}(1+x)\ln^2x + 6(1+x)\Li{2}{1-x},
\\ \label{PR}
&& P^{(0)}_{e\gamma}\otimes P^{(0)}_{\gamma e}
\otimes P^{(0)}_{ee}(x) = (1+x)\biggl(4\ln(1-x)\ln x
- \ln^2x + 4\Li{2}{1-x} \biggr)
\nonumber \\ && \qquad
+ \frac{2}{3}(3x + 4x^2)\ln x
+ \frac{2}{3}\biggl( \frac{4}{x} + 3 - 3x - 4x^2 \biggr)\ln(1-x)
- \frac{23}{6}(1-x),
\ea
where
\ba
\zeta_n \equiv \sum_{k=1}^{\infty}\frac{1}{k^n}\, ,\qquad
\zeta_2 = \frac{\pi^2}{6}\, ,\qquad
\Li{2}{x} \equiv - \int\limits_{0}^{x}\dd y\;\frac{\ln(1-y)}{y}\, .
\ea

By convolution of the fragmentation
functions [Eqs.~(\ref{De}) and (\ref{Dg})]
with the differential distributions~(\ref{dhGj}), we
receive higher order corrections to the electron spectrum,
which will be presented in Sect.~\ref{Sec:Results}.
In the results we can fix (see discussion in Ref.~\cite{Arbuzov:2002cn})
the factorization scale
\ba
\mu_f = m_{\mu}
\ea
and get $L_f\to L$.

\section{\label{Sec:Exp}Exponentiation}

Looking at the end point of the energy spectrum $(x\to 1)$ of
unpolarized muon decay,
one can recognize that the first order correction becomes there negative
and very large, making the result senseless. An extensive
discussion of the phenomenon can be found
in Refs.~\cite{Kinoshita:1959ru,Marciano:1975kw}.
The divergence is a clear signal to look beyond the first order
approximation. In fact, the Yennie--Frautschi--Suura
theorem~\cite{Yennie:1961ad}
allows to make a re--summation of the dangerous terms
and to convert them into a definitely positive exponent.
The exponentiation procedure is not unique, it permits to
involve also some terms convergent at $x\to 1$.
In our case, the exponentiation is allowed to add
several terms of the following types:
$\order{\alpha^2L^0}$, $\order{\alpha^nL^m}$
with $n\geq 3,\ 0\leq m < n$, and
$\order{\alpha^nL^n}$, $n\geq 4$.

Let us consider two ways of exponentiation. In the first case
one starts from the corrected cross section and tries to
perform a re--summation of the known terms, which are
divergent at $x\to 1$, by converting them into an exponent:
\ba \label{exp:ah}
\frac{F(x)}{f_0(x)}\bigg|_{x\to 1} &\approx& 1
+ \frac{\alpha}{\pi}(L-2)\ln(1-x)
+ \ldots \ \longrightarrow \
\mathrm{exp}\left\{\frac{\alpha}{\pi}(L-2)\ln(1-x)
\right\}.
\ea
This is a kind of the so--called {\it ad hoc} exponentiation.
The effect (see Table~\ref{table:1}) of this approach can be represented
by the relative contribution of {\it new} terms
generated by the exponent,
\ba \label{delta:ah}
\delta^{\mathrm{exp}}_{\mathrm{a.h.}}(x)
&=& \mathrm{exp}\left\{\frac{\alpha}{\pi}(L-2)\ln(1-x)\right\}
- 1 - \frac{\alpha}{\pi}(L-2)\ln(1-x)
\nonumber \\
&-& \frac{1}{2}\biggl(\frac{\alpha}{\pi}\biggr)^2
\bigl( L^2 - 4L \bigr)\ln^2(1-x)
- \frac{1}{6}\biggl(\frac{\alpha}{\pi}L\ln(1-x)\biggr)^3.
\ea
The most significant term above is
of the order $\alpha^2\ln^2(1-x)$ and gives a numerically
important contribution at large $x$. Note that all the subtracted terms
do appear in the perturbative results.

The next step should be to check that in higher orders the exponent
doesn't contradict the known (or anticipated) results.
The above procedure in the case of muon decay can be
criticized~\cite{Roos:1971mj}, because the higher order
leading logarithmic terms
represent a mass singularity: one can not guarantee that all the large
logarithms, coming from $\delta^{\mathrm{exp}}_{\mathrm{a.h.}}(x)$,
disappear after the integration over the energy fraction.
Nevertheless, the {\it ad hoc} exponentiation is not supposed to produce
a complete result. The region of its application is limited:
it deals with the terms, which are the most
important in the end of the spectrum.

There is another way of exponentiation, which avoids the problem
of improper mass singularities. One can use the exponentiated
representation of the electron structure (fragmentation) function
suggested in Ref.~\cite{Kuraev:1985hb}, which obeys the proper
normalization:
\ba \label{normd}
\int\limits_0^1\DD^{(\gamma)}_{\mathrm{exp}}(x)\dd x = 1.
\ea
The exponentiated structure functions are based on the exact solutions
of the QED evolution equations in the limiting case of soft radiation.
For computations I used an advanced formula from
Ref.~\cite{Cacciari:1992pz},
where I substituted $(L-1)$, which was natural for $e^+e^-$ annihilation,
by $(L-2)$. This substitution has not been obvious from
the beginning, but it is clearly seen in the above {\it ad hoc}
exponentiation. The usual $(L-1)$ factor corresponds
to soft radiation off the electron, while the additional $(-1)$ is due
to radiation off the muon. In fact one can introduce the muon structure
function into the master equation~(\ref{master}), as discussed in
Ref.~\cite{Cacciari:2001cw}. The muon structure function does not give
any large logarithms in our case. But still it can be used to describe
the contribution of soft photon radiation off the muon. And the corresponding
factor is just that $(-1)$ instead of the usual $(L-1)$. We see, that it
is easy to make a mistake in exponentiation of soft gluons (photons)
in decay processes by forgetting about the radiation off the decaying
particle.

The relevant for us electron structure function~\cite{Cacciari:1992pz} is
taken within the leading logarithm approximation for pure
photonic corrections with terms
up to $\order{\alpha^3L^3}$ and supplied with
exponentiation of some terms in higher orders.
Convolution with the Born--level functions gives
\ba \label{delta:SF}
F_{\mathrm{SF}}^{\mathrm{exp}}(x) =
\DD^{(\gamma)}_{\mathrm{exp}} \otimes f_0(x), \qquad
G_{\mathrm{SF}}^{\mathrm{exp}}(x) =
\DD^{(\gamma)}_{\mathrm{exp}} \otimes g_0(x).
\ea
A subtraction of the known terms in the lower orders $(n\leq 3)$
of the perturbative
expansion, similar to Eq.~(\ref{delta:ah}), is used to receive the value of
the relative contribution $\delta^{\mathrm{exp}}_{\mathrm{SF}}(x)$.
The latter contains also terms of the order $\order{\alpha^2L^0}$,
which are not singular at $x\to 1$. This is due to the fact
that soft radiation
is allowed not only at the end of the energy spectrum, but in any other point
in $x$. The resulting effect is spread all over the spectrum and appears to
be visible at the two ends $x\to 0$ and $x\to 1$.
Numerical results (see Table~\ref{table:1}) of the
two approaches are close to each other in the large-$x$ region (where an
analytical agreement between the approaches can be observed as well).
Contrary to the exponentiation of soft gluons in QCD~\cite{Cacciari:2001cw},
our procedure doesn't suffer from the renormalization scheme (and scale)
dependence and from non--perturbative effects.

Simultaneous exponentiation of photonic and pair corrections can be
constructed as well. But it was criticized~\cite{Arbuzov:rt}, since soft
pairs (contrary to soft photons) have a non--zero production
threshold, which can't be taken into account by exponentiation properly.

\section{\label{Sec:Results}Results}

To the best of our present knowledge, we can write now function
$F(x)$ from the master formula~(\ref{general}) as follows:
\ba \label{Ffinal}
F(x) &=& f_{\mathrm{Born}}(x) + \frac{\alpha}{2\pi}f_1(x)
+ \biggl(\frac{\alpha}{2\pi}\biggr)^2\biggl\{
\biggl[ f_2^{(0,\gamma)}(x) + \frac{2}{3}f_2^{(0,\NS)}(x)
+ f_2^{(0,\SS)}(x)\biggr] \frac{L^2}{2}
\nonumber \\
&+& \biggl[ f_2^{(1,\gamma)}(x) + f_2^{(1,\NS)}(x)
+ f_2^{(1,\SS)}(x) + f_2^{(1,\mathrm{int})}(x) \biggr] L
\biggr\}
\nonumber \\
&+& \biggl(\frac{\alpha}{2\pi}\biggr)^3\biggl[
f_3^{(0,\gamma)}(x) + f_3^{(0,\NS)}(x)
+ f_3^{(0,\SS)}(x) \biggr] \frac{L^3}{6}
+ \Delta f_{\mathrm{exp}}(x)
\nonumber \\
&+& \order{\alpha^2L^0,\alpha^3L^2,\alpha^4L^4}.
\ea
Function $G(x)$ takes the same form with the substitution
$f_i\rightarrow g_i$. Effects due to virtual hadronic,
$\mu^+\mu^-$, and $\tau^+\tau^-$ pairs~\cite{Davydychev:2001ee}
are not shown explicitly, since they are of the order $\order{\alpha^2L^0}$.
The effect of exponentiation is given by
\ba
\Delta f_{\mathrm{exp}}(x) = f_0(x)\delta^{\mathrm{exp}}_{\mathrm{SF}}(x),
\qquad
\Delta g_{\mathrm{exp}}(x) = g_0(x)\delta^{\mathrm{exp}}_{\mathrm{SF}}(x).
\ea

\subsection{Analytical Results}

In the second order for the NLL corrections to
the anisotropic part of the electron energy distribution, we have
\ba \label{g2nllg}
g_2^{(1,\gamma)}(x) &=&  4x^2(1-2x)\biggl(
\Li{3}{1-x} + \Sot{1-x} - 2\Li{2}{1-x}\ln(1-x)
\nonumber \\
&+& \ln x\ln^2(1-x) - 3\ln^2x\ln(1-x)
+ \ln^3x - \zeta_2\ln x + \frac{3}{2}\zeta_3 \biggr)
\nonumber \\
&+& \biggl( \frac{14}{3} - \frac{8}{3x} - 6x + 24x^2
- \frac{92}{3}x^3 \biggr)\Li{2}{1-x}
+ \biggl(  6x - 5 - \frac{86}{3}x^2 \biggr)\ln x \ln(1-x)
\nonumber \\
&+& \biggl( 8 - \frac{8}{3x} - 12x + \frac{20}{3}x^2 + 8x^3 \biggr)\ln^2(1-x)
+ \biggl( \frac{5}{12} + 18x^2 - \frac{70}{3}x^3 \biggr)\ln^2x
\nonumber \\
&+& \biggl( - \frac{13}{3} + \frac{37}{3}x + \frac{50}{3}x^2
- \frac{32}{3}x^3 \biggr)\ln(1-x)
+ \biggl( \frac{25}{12} - \frac{59}{6}x + 6x^2 + \frac{32}{9}x^3 \biggr)\ln x
\nonumber \\
&+& \biggl( - 8 + \frac{8}{3x} + 12x - \frac{29}{3}x^2 - 2x^3 \biggr)\zeta_2
+ \frac{817}{216} - \frac{91}{12}x + \frac{62}{3}x^2 - \frac{607}{54}x^3,
\\ \label{g2nllns}
g_{2}^{(1,\NS)}(x) &=&
4x^2(1-2x)\biggl( - \Li{2}{1-x} - \frac{1}{3}\ln x\ln(1-x)
+ \frac{1}{3}\ln^2(1-x)
\nonumber \\
&-& \frac{1}{2}\ln^2x - \frac{1}{3}\zeta_2 \biggr)
+ \biggl( \frac{22}{9} - \frac{8}{9x} - 4x - 6x^2 + 12x^3 \biggr)\ln(1-x)
\nonumber \\
&+& \biggl( - \frac{1}{9} + \frac{8}{9}x^2 - \frac{76}{9}x^3 \biggr)\ln x
- \frac{7}{18} + \frac{5}{3}x + \frac{86}{9}x^2 - \frac{20}{3}x^3,
\\ \label{g2nlls}
g_2^{(1,\SS)}(x) &=&
\biggl( \Li{2}{1-x} + \ln x\ln(1-x) \biggr)
\biggl( \frac{4}{3}x^2 - \frac{1}{3} \biggr)
+ \biggl( \frac{4}{3}x^2 - \frac{1}{2} \biggr)\ln^2x
\nonumber \\
&+& \biggl( - \frac{1}{9} - \frac{2}{9x} + x + \frac{2}{9}x^2
- \frac{8}{9}x^3 \biggr)\ln(1-x)
+ \biggl( \frac{5}{9} - \frac{4}{9x} + \frac{5}{2}x
+ \frac{5}{9}x^2 \biggr)\ln x
\nonumber \\
&+& \frac{1}{3x} + \frac{4}{3} - \frac{7}{18}x - \frac{43}{18}x^2
+ \frac{10}{9}x^3,
\\ \label{g2nlli}
g_2^{(1,\mathrm{int})}(x) &=&
4x^2(1-2x)\biggl( \Li{3}{1-x} - 2\Sot{1-x} - \Li{2}{1-x}\ln x \biggr)
\nonumber \\
&+& \biggl( - \frac{1}{3} - 14x^2 + \frac{52}{3}x^3 \biggr)\Li{2}{1-x}
+ \biggl(  - 3x^2 + \frac{26}{3}x^3 \biggr)\ln^2 x
\nonumber \\
&+& \biggl(  \frac{1}{3} + \frac{1}{3}x - \frac{28}{3}x^2 \biggr)\ln x
+ \frac{10}{9} - \frac{1}{3}x - \frac{37}{3}x^2 + \frac{104}{9}x^3.
\ea
The polylogarithm functions are defined as
\ba
\Li{3}{x} \equiv \int\limits_{0}^{x}\dd y\;\frac{\Li{2}{y}}{y}\, , \qquad
\Sot{x} \equiv \frac{1}{2}\int\limits_{0}^{x}
\dd y\;\frac{\ln^2(1-y)}{y}\, .
\ea
The $\order{\alpha^2L^2}$ corrections $f_{2}^{(0,j)}(x)$ and
$g_{2}^{(0,j)}(x)$ $(j=\gamma,\, \NS,\, \SS)$ can be found in
Ref.~\cite{Arbuzov:2002pp}.
Explicit expressions for the second order next--to--leading
corrections to the isotropic part of the spectrum
$(f_{2}^{(1,i)}(x),\ i=\gamma,\, \NS,\, \SS,\, \mathrm{int})$
are given in Ref.~\cite{Arbuzov:2002cn}.

The third order LL photonic contributions read
\ba \label{f3g}
f_{3}^{(0,\gamma)}(x) &=& 8x^2(3-2x)\Psi(x)
+ \bigl( 10 + 24x - 48x^2 + 32x^3 \bigr)\ln^2(1-x)
\nonumber \\
&+& \biggl( \frac{5}{12} + x - 8x^2 + 16x^3 \biggr)\ln^2x
+ \bigl( - 5 - 12x + 48x^2 - 64x^3\bigr)\ln x\ln(1-x)
\nonumber \\
&+& \bigl(5+12x-32x^3\bigr)\Li{2}{1-x}
+ \bigl(-10-24x+48x^2-32x^3\bigr)\zeta_2
\nonumber \\
&+& \biggl( -\frac{13}{18} - \frac{21}{2}x + \frac{64}{3}x^3 \biggr)\ln x
+ \biggl( \frac{11}{6} + 17x + 16x^2 - \frac{64}{3}x^3 \biggr)\ln(1-x)
\nonumber \\
&+& \frac{569}{216} + \frac{4}{3}x - \frac{16}{3}x^2 + \frac{128}{27}x^3,
\\ \label{g3g}
g_{3}^{(0,\gamma)}(x) &=& 8x^2(1-2x)\Psi(x)
+ \bigl( - 2 - 48x^2 + 32x^3 \bigr)\ln^2(1-x)
\nonumber \\
&+& \biggl( - \frac{1}{12} - 8x^2 + 16x^3 \biggr)\ln^2x
+ \bigl( 1 + 48x^2 - 64x^3\bigr)\ln x\ln(1-x)
\nonumber \\
&+& \bigl(-1-32x^3\bigr)\Li{2}{1-x}
+ \bigl(2+48x^2-32x^3\bigr)\zeta_2
\nonumber \\
&+& \biggl( \frac{5}{18} + \frac{5}{2}x + \frac{64}{3}x^3 \biggr)\ln x
+ \biggl( - \frac{7}{6} - 7x + 16x^2 - \frac{64}{3}x^3 \biggr)\ln(1-x)
\nonumber \\
&-& \frac{133}{216} - \frac{13}{6}x - \frac{16}{3}x^2 + \frac{128}{27}x^3,
\\ \nonumber
\Psi(x) &\equiv& 3\Li{3}{1-x} - 2\Sot{1-x}
+ \ln^3(1-x) - \frac{1}{6}\ln^3x + \frac{3}{2}\ln^2x \ln(1-x)
\\ \nonumber
&-& 3 \ln x\ln^2(1-x) - 3\Li{2}{1-x}\ln(1-x) + 2\zeta_3
- 3\zeta_2\ln\frac{1-x}{x}\, .
\ea
And the third order LL pair corrections\footnote{Strictly speaking,
we have here pair and photonic corrections simultaneously.} are
\ba \label{f3ns}
f_3^{(0,\NS)}(x) &=& 8x^2(3-2x) \Phi(x)
+ \biggl( \frac{20}{3} + 16x - \frac{80}{3}x^2
+ \frac{160}{9}x^3 \biggr)\ln(1-x)
\nonumber \\
&+& \biggl( - \frac{5}{3} - 4x + \frac{32}{3}x^2
- \frac{160}{9}x^3 \biggr)\ln x
+ \frac{73}{54} + \frac{67}{9}x + \frac{16}{9}x^2 - \frac{128}{27}x^3,
\\ \label{g3ns}
g_3^{(0,\NS)}(x) &=&
8x^2(1-2x) \Phi(x)
+ \biggl( - \frac{4}{3} - \frac{272}{9}x^2 + \frac{160}{9}x^3 \biggr)\ln(1-x)
\nonumber \\
&+& \biggl( \frac{1}{3} + \frac{128}{9}x^2 - \frac{160}{9}x^3 \biggr)\ln x
- \frac{29}{54} - \frac{7}{3}x + \frac{16}{9}x^2 -\frac{128}{27}x^3,
\\ \label{f3s}
f_3^{(0,\SS)}(x) &=& \biggl( \frac{5}{3} + 4x + 4x^2 \biggr)\biggl(
4\Li{2}{1-x} + 4\ln x\ln(1-x) - \ln^2x \biggr) - 4x^2\ln^2x
\nonumber \\
&+&  \biggl( \frac{68}{9} + \frac{8}{3x} + 12x - \frac{56}{3}x^2
- \frac{32}{9}x^3 \biggr)\ln(1-x)
+ \biggl( - \frac{29}{9} - \frac{14}{3}x + 16x^2
\nonumber \\
&+& \frac{32}{9}x^3 \biggr)\ln x
- \frac{287}{27} - \frac{4}{9x} - \frac{13}{9}x
+ \frac{86}{9}x^2 + \frac{80}{27}x^3,
\\ \label{g3s}
g_3^{(0,\SS)}(x) &=& \biggl( \frac{4}{3}x^2 - \frac{1}{3} \biggr)
\biggl( 4\Li{2}{1-x} + 4\ln x\ln(1-x) - \ln^2x \biggr)
- \frac{4}{3}x^2\ln^2x
\nonumber \\
&+& \biggl(  - \frac{4}{9} - \frac{8}{9x} + 4x
+ \frac{8}{9}x^2  - \frac{32}{9}x^3 \biggr)\ln(1-x)
\nonumber \\
&+& \biggl( \frac{1}{9} - 2x - \frac{16}{9}x^2
+ \frac{32}{9}x^3 \biggr)\ln x
+ \frac{31}{27} + \frac{4}{27x} - \frac{35}{9}x
- \frac{10}{27}x^2 + \frac{80}{27}x^3,
\\ \nonumber
\Phi(x) &\equiv& \frac{1}{2}\ln^2 x + \ln^2(1-x)
- 2\ln x\ln(1-x) - \Li{2}{1-x} - \zeta_2.
\ea

Functions
$g_2^{(1,\SS)}$ and $g_2^{(1,\mathrm{int})}$, shown above, as well as
$f_2^{(1,\SS)}$ and $f_2^{(1,\mathrm{int})}$, which are given
in Ref.~\cite{Arbuzov:2002cn}, coincide with
the results of my calculations starting directly from Feynman
diagrams and using methods described in Ref.~\cite{Arbuzov:1995cn}.

\subsection{Cancellation of Mass Singularities}

An important check of the results is to demonstrate the cancellation
of mass singularities in the total decay width.
The cancellation of the mass singularities in the LL contributions
due to photons and non--singlet pairs is rather trivial:
\ba
\int\limits_{0}^{1} \dd x\; f_{n}^{(0,j)}(x) =
\int\limits_{0}^{1} \dd x\; g_{n}^{(0,j)}(x) = 0, \qquad
j = \NS,\ \gamma, \qquad n = 1,2,3,\ldots
\ea
It is guaranteed by the normalization conditions of the corresponding
LL fragmentation functions.

Now we should note that a naive integration of the electron spectrum
gives rather the counting rate of electrons than the total muon
decay width, since the number of the final state electrons can exceed
the number of decaying muons because of real $e^+e^-$ pair emission.
In other words, we should avoid the double counting of electrons in
the contributions due to real pair emission. For this purpose,
we can keep the non--singlet pair contributions and drop the
singlet ones (see their definition on page~\pageref{def:sns}).

Functions $f_2^{(1,\mathrm{int})}$ and $g_2^{(1,\mathrm{int})}$
contain the double counting too. To resolve this
problem we can use the splitting function~\cite{Curci:1980uw,Altarelli:1979kv}
\ba
P^{(1,\mathrm{int})}_{\bar{e}e}(x) &=&
2\frac{1+x^2}{1+x}\biggl( -2\Li{2}{-x} - 2\ln x\ln(1+x) + \frac{1}{2}\ln^2x
- \zeta_2 \biggr)
\nonumber \\
&+& 2(1+x)\ln x + 4(1-x),
\ea
which describes the transition of an electron into a positron
in the relevant set of Feynman diagrams. The corresponding
contribution can be constructed by convolution with the lowest order
functions $f_0$ and $g_0$ integrated over the positron energy
fraction. One can see now that in the interference contribution
the number of electrons is really twice as large as
the number of positrons:
\ba \label{docof}
\int\limits_{0}^{1}\dd x f_2^{(1,\mathrm{int})}(x)
= 2\int\limits_{0}^{1}\dd x
P^{(1,\mathrm{int})}_{\bar{e}e}\otimes f_0(x) =
\frac{13}{4} - 3\zeta_2 + 2\zeta_3.
\ea
By resolving the problem of double counting in the
contribution of the S-NS pair interference, we arrive
to the cancellation of mass singularities in
the following form:
\ba \label{iintf}
&& \int\limits_{0}^{1}\dd x\;
( f_2^{(1,\gamma)}(x) + f_2^{(1,\mathrm{int})}(x)
- P^{(1,\mathrm{int})}_{\bar{e}e}\otimes f_0(x) )
\nonumber \\ && \qquad
= \int\limits_{0}^{1}\dd x\;
( f_2^{(1,\gamma)}(x) + \frac{1}{2}f_2^{(1,\mathrm{int})}(x) ) = 0.
\ea

Let us look now at the integral of the second order NLL
non--singlet pair correction. It is known~\cite{vanRitbergen:1998yd},
that the integrated contribution of this correction contains
large logarithms due to the running of the coupling constant:
\ba
\biggl(\frac{\alpha}{2\pi}\biggr)^2L\int\limits_{0}^{1}\dd x\;
f_2^{(1,\NS)}(x) &=&
\frac{\Delta\alpha(m_\mu)}{2\pi} \biggl[
\int\limits_{0}^{1}\dd x\; f_1(x) \biggr] \bigg|_{m_e\to 0},
\ea
where
\ba
\Delta\alpha(m_\mu) = \frac{\alpha^2}{3\pi}L
= \alpha(m_\mu) - \alpha + \order{\alpha^2L^0}.
\ea
To demonstrate that using our results, note first that the relevant
function consists of two parts:
\ba
f_2^{(1,\NS)}(x) = P^{(1,\NS)}_{ee}\otimes f_0(x)
+ \frac{2}{3}\hat{f}_1(x).
\ea
One can check that
\ba
\int\limits_{0}^{1}\dd x\;
P^{(1,\NS)}_{ee}\otimes f_0(x) = 0.
\ea
It remains now to recognize that
\ba
\biggl[ \int\limits_{0}^{1}\dd x\; f_1(x) \biggr] \bigg|_{m_e\to 0}
= \int\limits_{0}^{1}\dd x\; \hat{f}_1(x),
\ea
which can be verified easily. Thus we checked successfully
an important property of our analytical results.

The anisotropic contributions to the decay spectrum
can be treated in the same way.
They don't contribute to the total decay width at all,
because they vanish after the integration over the angle.
Nevertheless, the cancellation of mass
singularities can be observed also in the forward--backward asymmetry
of the decay, which is not affected by isotropic functions
on the contrary.
In particular, we have an equality analogous to Eq.~(\ref{docof}):
\ba \label{docog}
\int\limits_{0}^{1}\dd x g_2^{(1,\mathrm{int})}(x)
= 2\int\limits_{0}^{1}\dd x
P^{(1,\mathrm{int})}_{\bar{e}e}\otimes g_0(x) =
- \frac{13}{12} + \zeta_2 - \frac{2}{3}\zeta_3.
\ea

\subsection{Numerical Results}

Now we can make numerical estimates of the effects due to higher order
radiative corrections. We should confront them with the $1\cdot 10^{-4}$ precision level
of the \TWIST experiment. In a typical experiment on muon decays one
has almost 100\% longitudinal polarization of muons ($P_{\mu}=1$),
since the muons are coming polarized from pion decays.
In order to extract information
about the Michel parameters $\xi$ and $\delta$ one should study carefully
the angular distribution of the electrons. But as can be seen from the
analytical formulae, the dependence on the angle is rather simple and smooth.
For this reason we restrict our numerical illustrations to the consideration
of $x$-dependence at several fixed values of the angle.

Numerical results for the pure photonic corrections are presented in
Table~\ref{table:1},
where we give the values of different contributions normalized by
the Born distribution in the following way:
\ba \label{delta_g}
&& \delta_1 = \frac{\alpha}{2\pi}\;\frac{f_1(x) \pm cP_{\mu}g_1(x)}
{f_{\mathrm{Born}}(x) \pm cP_{\mu}g_{\mathrm{Born}}(x)}\, ,
\nonumber \\
&& \delta_n^{(0,\gamma)} = \frac{L^n}{n!}
\biggl(\frac{\alpha}{2\pi}\biggr)^n\frac{f_n^{(0,\gamma)}(x)
\pm cP_{\mu}g_n^{(0,\gamma)}(x)}
{f_{\mathrm{Born}}(x) \pm cP_{\mu}g_{\mathrm{Born}}(x)}\, ,\qquad
n = 1,2,3\; ,
\nonumber \\
&& \delta_2^{(1,\gamma)} = L\biggl(\frac{\alpha}{2\pi}\biggr)^2
\frac{f_2^{(1,\gamma)}(x) \pm cP_{\mu}g_2^{(1,\gamma)}(x)}
{f_{\mathrm{Born}}(x) \pm cP_{\mu}g_{\mathrm{Born}}(x)}\, .
\ea
\TABLE[ht]{
\label{table:1}
\begin{tabular}{|r|r|r|r|r|r|r|r|} \hline
\multicolumn{1}{|c|}{$x$} &
\multicolumn{1}{c|}{$10^4\delta_1$} &
\multicolumn{1}{c|}{$10^4\delta_1^{(0,\gamma)}$} &
\multicolumn{1}{c|}{$10^4\delta_2^{(0,\gamma)}$} &
\multicolumn{1}{c|}{$10^4\delta_2^{(1,\gamma)}$} &
\multicolumn{1}{c|}{$10^4\delta_3^{(0,\gamma)}$} &
\multicolumn{1}{c|}{$10^4\delta^{\mathrm{exp}}_{\mathrm{a.h.}}$} &
\multicolumn{1}{c|}{$10^4\delta^{\mathrm{exp}}_{\mathrm{SF}}$}
\\ \hline
0.05 &   4590.1 &  10325.0 &  184.96 &$-$247.63 & $-$0.43 & 0.00 &    6.90 \\
0.1  &   1715.1 &   3257.7 &   33.18 & $-$37.79 & $-$0.35 & 0.00 &    1.39 \\
0.2  &    674.0 &   1106.2 & $-$1.28 &     0.34 & $-$0.15 & 0.01 &    0.05 \\
0.3  &    364.0 &    549.1 & $-$6.58 &     4.48 & $-$0.05 & 0.01 & $-$0.20 \\
0.5  &     64.1 &     82.6 & $-$6.50 &     3.73 &    0.05 & 0.06 & $-$0.26 \\
0.7  & $-$160.3 & $-$214.9 & $-$1.28 &     0.70 &    0.07 & 0.18 & $-$0.09 \\
0.9  & $-$470.3 & $-$592.1 &   15.19 &  $-$5.97 & $-$0.14 & 0.72 &    0.62 \\
0.99 & $-$971.9 &$-$1198.8 &   69.84 & $-$26.08 & $-$1.95 & 3.47 &    3.70 \\
0.999&$-$1439.8 &$-$1772.5 &  155.10 & $-$57.86 & $-$6.64 & 9.17 &    9.75 \\
\hline
\end{tabular}
\caption{Photonic corrections to $\mu^-$ decay spectrum
versus $x$ for $c=1$ and $P_{\mu}=1$.}
}
One can see that the first order LL correction $\delta_1^{(0,\gamma)}$
(look for $f_1^{(0,\gamma)}(x)$ and $g_1^{(0,\gamma)}(x)$ in
Ref.~\cite{Arbuzov:2002pp})
provides the bulk of the effect, especially in the region of intermediate
and large $x$-values. Convergence of the corresponding series in $L$
in the second order corrections doesn't look so good: the NLL contribution
is only about two times less than the LL one.

There is a trick, which allows to make a better approximation in
the region of small $x$. Looking closely at the argument of
the large logarithm during the actual calculations of integrals
over the phase space of real photons, one can notice that it is
rather $x^2m_\mu^2/m_e^2$, than simply $m_\mu^2/m_e^2$. So the
modification $L \to L + 2\ln x$ can be done in our formulae.
This will move some terms from the sub--leading corrections
into the leading ones. I checked that the trick does really help
to improve the agreement in the first order between
$\delta_1^{(0,\gamma)}$ and the full $\delta_1$.
But, as far as the \TWIST experiment is
interested in the region $x \geq 0.3$ (the event distribution
is peaked at large $x$-values in any case), I don't apply the modification here.

Let us define the relative contributions of pair corrections as
\ba
&& \delta_2^{(0,e\bar{e})} = \frac{L^2}{6}\biggl(\frac{\alpha}{2\pi}\biggr)^2
\frac{ 2f_2^{(0,\NS)}(x) + 3f_2^{(0,\SS)}(x)
\pm cP_{\mu}(2g_2^{(0,\NS)}(x)+3g_2^{(0,\SS)}(x))}
{f_{\mathrm{Born}}(x) \pm cP_{\mu}g_{\mathrm{Born}}}\, ,
\nonumber \\
&& \delta_2^{(1,e\bar{e})} = L
\biggl(\frac{\alpha}{2\pi}\biggr)^2
\sum\limits_{i=\NS,\SS,\mathrm{int}}^{}
\frac{f_2^{(1,i)}(x) \pm cP_{\mu}g_2^{(1,i)}(x)}
{f_{\mathrm{Born}}(x) \pm cP_{\mu}g_{\mathrm{Born}}(x)}\, ,
\nonumber \\
&& \delta_3^{(0,e\bar{e})} = \frac{L^3}{6}
\biggl(\frac{\alpha}{2\pi}\biggr)^3
\sum\limits_{j=\NS,\SS}^{}
\frac{f_3^{(0,j)}(x) \pm cP_{\mu}g_3^{(0,j)}(x)}
{f_{\mathrm{Born}}(x) \pm cP_{\mu}g_{\mathrm{Born}}(x)}\, .
\ea
For two values of $c$, they are given in Table~\ref{table:2}.
\TABLE[ht]{
\label{table:2}
\begin{tabular}{|r|r|r|r||r|r|r|} \hline
\multicolumn{1}{|c|}{} &
\multicolumn{3}{c||}{$c=1$} &
\multicolumn{3}{c|}{$c=-1$} \\ \hline
\multicolumn{1}{|c|}{$x$} &
\multicolumn{1}{c|}{$10^4\delta_2^{(0,e\bar{e})}$} &
\multicolumn{1}{c|}{$10^4\delta_2^{(1,e\bar{e})}$} &
\multicolumn{1}{c||}{$10^4\delta_3^{(0,e\bar{e})}$} &
\multicolumn{1}{c|}{$10^4\delta_2^{(0,e\bar{e})}$} &
\multicolumn{1}{c|}{$10^4\delta_2^{(1,e\bar{e})}$} &
\multicolumn{1}{c|}{$10^4\delta_3^{(0,e\bar{e})}$}
\\ \hline
0.05 & 548.14 &$-$661.94& $-$2.19 &2132.14 &$-$2848.25& $-$9.12 \\
0.1  &  66.02 & $-$68.36& $-$0.18 & 241.62 &$-$274.34 & $-$0.76 \\
0.2  &   9.55 & $-$9.99 & $-$0.05 &  29.25 &$-$29.68  & $-$0.04 \\
0.3  &   3.43 & $-$3.58 & $-$0.06 &   9.59 & $-$9.77  & $-$0.05 \\
0.5  &   0.49 & $-$0.71 & $-$0.06 &   2.54 & $-$2.99  & $-$0.08 \\
0.7  &$-$0.86 &    0.76 & $-$0.02 &   0.46 & $-$1.03  & $-$0.08 \\
0.9  &$-$2.44 &    2.95 &    0.05 &$-$1.36 &    0.99  &    0.00 \\
0.99 &$-$4.95 &    8.24 & $-$0.73 &$-$3.92 &    5.47  &    0.33 \\
0.999&$-$7.32 &   15.41 & $-$18.18&$-$6.29 &   11.76  &    0.92 \\
\hline
\end{tabular}
\caption{Pair corrections to $\mu^-$ decay spectrum
{\it versus} $x$ for $c=\pm 1$ and $P_{\mu}=1$.}
}
One can see that the next--to--leading pair corrections
have the same order of magnitude as the leading ones.
This feature has been observed earlier in pair corrections
to other processes~\cite{Arbuzov:rt}.
The functions, which describe the LL pair effect, have
numerically small coefficients and are less divergent
(at $x\to 1$ and $x\to 0$)
than the NLL ones. This means that the calculations of the
non--logarithmic terms in the second order pair corrections
is desirable, although the pair corrections are typically
less than the photonic ones (at the same order in $\alpha$).

The combined effect of the discussed above higher order corrections
to $\mu^-$ decay spectrum is shown in Fig.~\ref{figure:1}:
\ba
\delta_{\mathrm{h.o.}} = \delta_2^{(0,\gamma)} + \delta_2^{(1,\gamma)}
+ \delta_3^{(0,\gamma)} + \delta_2^{(0,e\bar{e})}
+ \delta_2^{(1,e\bar{e})} + \delta_3^{(0,e\bar{e})}
+ \delta^{\mathrm{exp}}_{\mathrm{SF}}.
\ea
\FIGURE[ht]{
\epsfig{file=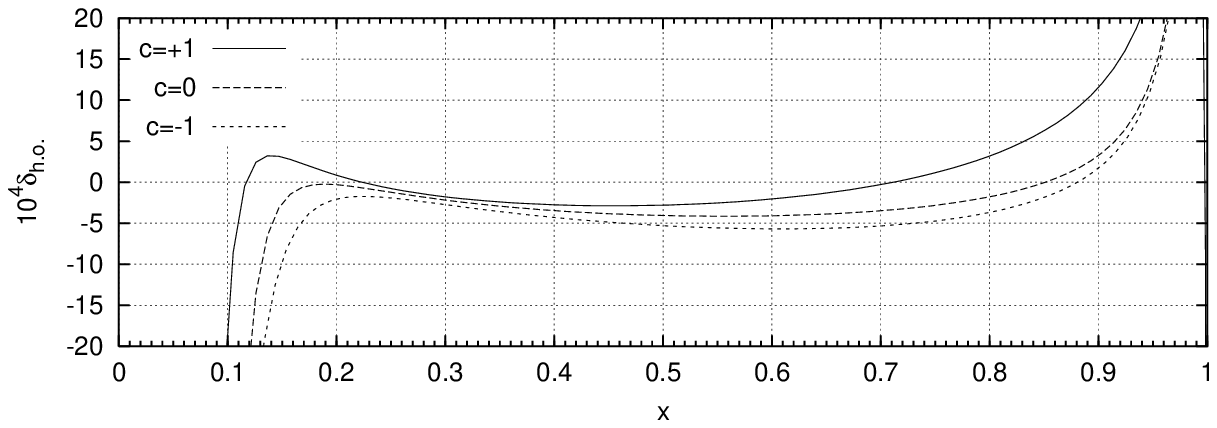,width=10cm,height=6cm}
\caption{The relative effect of higher order corrections {\it versus}
electron energy fraction for different angles.}
\label{figure:1}
}
One can see from the Figure that the typical effect is of the order
of $5\cdot 10^{-4}$ for the intermediate range of $x$, and the effect
becomes even larger at the two ends of the energy spectrum. So the
corrections under consideration are really important for
the modern and future experimental studies of the muon decay spectrum,
where elaborated technique and high statistics allow to reduce the
experimental errors to the level of $1\cdot 10^{-4}$ and better.

\section{\label{Sec:Con}Conclusions}

To estimate the theoretical uncertainty in the description
of the polarized muon decay spectrum by Eqs.~(\ref{general})
and (\ref{Ffinal}), we should consider the contributions,
which have been omitted in the present calculations. They are:
the $\order{\alpha^2}$ order terms, which are not enhanced by the
large logarithm; sub--leading contributions in the third order
$\order{\alpha^3L^m}$, $m\leq 2$; and all the leading and sub--leading
effects in the forth and higher orders $(\order{\alpha^nL^m}$ where
$n \geq 4$, $m\leq n)$ except those ones, which are taken
into account by exponentiation. The possible contribution to
the uncertainty from strong interactions is negligible in our case,
since it is suppressed at least by $(\alpha/\pi)^2$, and the lowest order
the contribution of hadronic virtual pairs was found in
Ref.~\cite{Davydychev:2001ee} to be small itself.

An estimate of the omitted contributions by a simple counting of
powers of the fine--structure constant and the large logarithm is
not very safe, because there could be some extra enhancement factors,
like numerically large constant coefficients or powers of $\ln(1-x)$
(the latter is partially taken into account by exponentiation).
I suggest to estimate the omitted terms
by a linear extrapolation of the known expansions in $\alpha$ and $L$.
Namely,
\ba \label{estimate}
\delta_2^{(2)} \sim \delta_2^{(1)}
\frac{\delta_2^{(1)}}{\delta_2^{(0)}}\, ,
\qquad
\delta_3^{(1)} \sim \delta_2^{(1)}\frac{\delta_3^{(0)}}{\delta_2^{(0)}}\, ,
\qquad
\delta_4^{(0)} \sim \delta_3^{(0)}\frac{\delta_3^{(0)}}{\delta_2^{(0)}}\, ,
\ea
where $\delta_2^{(2)}$ denotes the contribution of
the second order terms, which are not enhanced by any
large logarithm; $\delta_3^{(1)}$ is the third order next--to--leading
correction (which can be calculated using the fragmentation function
method described above); $\delta_4^{(0)}$ stands for the fourth
order LL effect.
This estimate of the theoretical error, can be applied to any
particular set of experimental conditions to derive the actual
uncertainty. In principle, the latter can depend
on various cuts and details of particle registration and event selection
(see discussion in Ref.~\cite{Arbuzov:2002pp}).
The approach~(\ref{estimate}) to estimate the uncertainty in the muon
decay spectrum description works well for the main part of the
kinematical domain. But, if one studies separately the extreme region
$x \ll 1$ (or $1-x \ll 1)$, a special investigation of the convergence
properties of our perturbative expansions in $\alpha$ and $L$
should be performed.
Evaluation of the uncertainties for any concrete experiment
can be done using the analytical results and applying
specific conditions of particle registration and data
fitting.

The new results of Ref.~\cite{Arbuzov:2002cn} and the present paper
reduce the theoretical uncertainty in the description of the polarized
muon decay spectrum. For the {\em quasi--realistic} experimental
setup described in Ref.~\cite{Arbuzov:2002pp}, we can obtain now
about 1.5 times better precision, so that, for instance, the
theoretical uncertainty for the Michel parameter $\rho$ becomes
$2\cdot 10^{-4}$ instead of $3\cdot 10^{-4}$ obtained in
Ref.~\cite{Arbuzov:2002pp}.
Nevertheless, this is still worse than the experimental
precision $1\cdot 10^{-4}$ planned at
\TWIST~\cite{Rodning:2001js,Quraan:2000vq}.
Assuming that a theoretical precision of about
one third (or less) of the experimental one would not spoil
results of an experiment, we see a challenge for further
investigations. First of all,
a calculation of the $\order{\alpha^2}$
contributions, which are not enhanced
by the large logarithm, is required to ameliorate
the theoretical precision. This calculation is difficult,
but possible by means of the standard methods.

The formulae for higher order corrections (with simple substitutions)
are valid also for the decays of $\tau$-lepton:
$\tau\to\mu\nu_{\tau}\bar{\nu}_{\mu}$ and
$\tau\to e\nu_{\tau}\bar{\nu}_{e}$.

\acknowledgments
This research was supported by the Natural Sciences
and Engineering Research Council of Canada.
I am grateful to A.~Czarnecki, A.~Gaponenko, and K.~Melnikov
for fruitful discussions.


\end{document}